\documentclass[showpacs,preprintnumbers,amsmath,amssymb,nofootinbib,floatfix]{revtex4}
\usepackage[dvips]{graphicx}% Include figure files
\usepackage{dcolumn}% Align table columns on decimal point
\usepackage{bm}% bold math

\def\Journal#1#2#3#4{{#1} {\bf #2}, #3 (#4)}
\def\AA{Astron. Astrophys. }

\def\APJ{Astrophys. J.}

\def\IJMPA{Int. J. Mod. Phys. A}

\def\JCAP{J. Cosmol. Astropart. Phys.}

\def\JCAP{JCAP}
\def\JHEP{JHEP}

\def\JETPUSSR{JETP (USSR)}
 
\def\MPLA{Mod. Phys. Lett. A}

\def\NPB{Nucl. Phys. B}

\def\NPBSUPPL{Nucl. Phys. B. Proc. Suppl.}

\def\PLB{Phys. Lett. B}

\def\PLBOLD{Phys. Lett.}

\def\PRL{Phys. Rev. Lett.}
\def\PRD{Phys. Rev. D}

\def\PTP{Prog. Theor. Phys.}
\def\RMP{Rev. Mod. Phys.}

\def\SCIENCE{Science}

\def\ZETP{Zh. Eksp. Teor. Piz.}

\begin{document}

%\preprint{TOKAI-HEP/TH-0505}

% \title{Majorana CP Violation from Dirac CP Violation}
 \title{Maximal CP Violation in Minimal Seesaw Model}
% Force line breaks with \\

\author{Teruyuki Kitabayashi}
\email{teruyuki@tokai-u.jp}

% \affiliation[Also at ]{Physics Department, XYZ University.}%Lines break automatically or can be forced with \\
\author{Masaki Yasu\`{e}}%
\email{yasue@keyaki.cc.u-tokai.ac.jp}
\affiliation{\vspace{5mm}%
\sl Department of Physics, Tokai University,\\
4-1-1 Kitakaname, Hiratsuka, Kanagawa 259-1292, Japan\\
}

\date{May, 2016}% It is always \today, today,

%%-------------------------------------------------
%% Abstract
%%-------------------------------------------------
\begin{abstract}
In the minimal seesaw model, we derive required constraints on Dirac neutrino masses inducing maximal CP violation in neutrino oscillations. If the maximal atmospheric neutrino mixing is further assumed, Dirac neutrino masses are uniquely determined to respect $\mu$-$\tau$ flavored CP symmetry for neutrinos.
\end{abstract}

\maketitle

% \linenumbers
%\section*{}

%% Main body
%%-------------------------------------------------
%%%%%%%%%%%%%%%%%%%%%%%%%%%%%%%%%%%%%%%%%
\section{\label{sec:1}Introduction}
%%%%%%%%%%%%%%%%%%%%%%%%%%%%%%%%%%%%%%%%%
Neutrino oscillations have been theoretically predicted \cite{PMNS} and experimentally observed as  atmospheric, solar, accelerator and reactor neutrino oscillations for more than a decade \cite{atmospheric,solarold,solar,reactor,accelerator,sin13}. Extensive analyses of the current experimental data on neutrino oscillations seem to suggest the presence of the Dirac CP violation in neutrino physics \cite{NuData}.  The Dirac CP violation is described by the CP-violating Dirac phase $\delta_{CP}$, which turns out to lie in the $1\sigma$-region of $\delta_{CP}/\pi = 1.13-1.64$ for the normal mass hierarchy (NH) or of $\delta_{CP}/\pi = 1.07-1.67$ for the inverted mass hierarchy (IH) \cite{NuData1}. There is another type of CP violation called Majorana CP violation. The relevant CP-violating phases are the Dirac phase and the Majorana phase \cite{CPViolationOrg}, which enter into the Pentecorvo-Maki-Nakagawa-Sakata mixing matrix $U_{PNMS}$ \cite{PMNS} that converts the mass eigenstates of neutrinos $\nu_{1,2,3}$ into the flavor neutrinos $\nu_{e,\mu,\tau}$.  Denoting the atmospheric neutrino mixing angle by $\theta_{23}$, the solar neutrino mixing angle by $\theta_{12}$  and the reactor neutrino mixing angle by $\theta_{13}$, the standard parametrization of $U_{PNMS}$ is given by the Particle Data Group (PDG) \cite{PDG} to be $U_{PDG}=U^{PDG}_\nu K^{PDG}$:
%%%%%%%%%%%%%%%%%%%%
\begin{eqnarray}
U_\nu^{PDG}&=&\left( {\begin{array}{*{20}{c}}
{{c_{12}}{c_{13}}}&{{s_{12}}{c_{13}}}&{{s_{13}}{e^{ - i{\delta _{CP}}}}}\\
{ - {{s_{12}{c_{23}}} - {c_{12}}{s_{23}}{s_{13}}{e^{i{\delta _{CP}}}}} }&{{c_{12}}{c_{23}} - {s_{12}}{s_{23}}{s_{13}}{e^{i{\delta _{CP}}}}}&{{s_{23}}{c_{13}}}\\
{{s_{12}}{s_{23}} - {c_{12}}{c_{23}}{s_{13}}{e^{i{\delta _{CP}}}}}&{ - {c_{12}}{{s_{23}} - {s_{12}}{c_{23}}{s_{13}}{e^{i{\delta _{CP}}}}} }&{{c_{23}}{c_{13}}}
\end{array}} \right),
\nonumber\\
{K^{PDG}} &=& \left( {\begin{array}{*{20}{c}}
{{e^{i{\phi _1}/2}}}&0&0\\
0&{{e^{i{\phi _2}/2}}}&0\\
0&0&{{e^{i{\phi _3}/2}}}
\end{array}} \right),
\label{Eq:U_PDG}
\end{eqnarray}
%%%%%%%%%%%%%%%%%%%%
for $c_{ij}=\cos\theta_{ij}$, $s_{ij}=\sin\theta_{ij}$ and similarly $t_{ij}=\tan\theta_{ij}$ ($i,j$=1,2,3), where $\phi_{1,2,3}$ stand for the Majorana phases, from which two independent combinations become the CP-violating Majorana phases.  

It is interesting to note that the experimentally allowed region of $\delta_{CP}$ includes $\delta_{CP} = 3\pi/2$ indicating maximal CP violation.  From the theoretical point of view, $\delta_{CP}$ arises from phases of flavor neutrino masses to be denoted by $M_{ij}$ ($i,j=e,\mu,\tau$). We have been advocating the following useful relation among $\delta_{CP}$ and $M_{ij}$ \cite{Useful1, Useful2}:
%%%%%%%%%%%%%%%%%%%%
\begin{equation}
\frac{{{M_{\tau \tau }} - {M_{\mu \mu }}}}{2}\sin 2{\theta _{23}} - {M_{\mu \tau }}\cos 2{\theta_{23}} = \tan {\theta_{13}}\left( {{M_{e\mu }}\cos {\theta_{23}} - {M_{e \tau }}\sin {\theta_{23}}} \right){e^{ - i{\delta _{CP}}}},
\label{Eq:M_nu for 23}
\end{equation}
%%%%%%%%%%%%%%%%%%%%
which is used to express $\theta_{23}$ in terms of $M_{ij}$.  The maximal CP violation can be induced if 
%%%%%%%%%%%%%%%%%%%%
\begin{equation}
M_{\tau \tau } - M_{\mu \mu } = {\rm imaginary},
\label{Eq:Maximal-CP1}
\end{equation}
%%%%%%%%%%%%%%%%%%%%
as well as 
%%%%%%%%%%%%%%%%%%%%
\begin{equation}
M_{\mu \tau } = {\rm imaginary},
\quad
M_{e\mu}\cos\theta_{23} - M_{e\tau}\sin\theta_{23}= {\rm real},
\label{Eq:Maximal-CP2}
\end{equation}
%%%%%%%%%%%%%%%%%%%%
for $\cos 2\theta_{23}\neq 0$, or 
%%%%%%%%%%%%%%%%%%%%
\begin{equation}
M_{e\mu}-\sigma M_{e\tau}= {\rm real},
\label{Eq:Maximal-CP2 and ATM}
\end{equation}
%%%%%%%%%%%%%%%%%%%%
for $\cos 2\theta_{23}= 0$ indicating the maximal atmospheric neutrino mixing, where $\sigma=\pm 1$ takes care of the sign of $\sin\theta_{23}$. 

%%%%%%%%%%%%%%%%%%%%
From the recent result of the Planck \cite{Planck2015arXiv}, the upper limit of the neutrino masses is given by $\sum m_\nu \le 0.17$ eV. On the other hand, the neutrino oscillation experiments measure $\Delta m_{31}^2 = m_3^2-m_1^2$ and $\Delta m_{32}^2 = m_3^2-m_2^2$. Choosing $\Delta m_{31}^2= 2.46\times 10^{-3}$ eV$^2 (\sim m_3^2)$ for NH with $m^2_3\gg m^2_2\gg m^2_1$ and $\Delta m_{32}^2 = -2.45\times 10^{-3}$ eV$^2 (\sim -m_2^2)$ for IH with $m^2_2> m^2_1 \gg m^2_3$ \cite{NuData}, we obtain that the heaviest neutrino mass, either $m_3$ or $m_2$, is approximately estimated to be $0.05$ eV.
%%%%%%%%%%%%%%%%%%%%
Why neutrinos are so light is a puzzling question to be solved. One of the promising theoretical ideas is the one based on the seesaw mechanism \cite{Seesaw}, which utilizes right-handed neutrinos.  The right-handed neutrinos can provide Dirac masses for flavor neutrinos and light flavor neutrinos can be generated if the right-handed neutrinos are very heavy.  Furthermore, CP violation in the early universe is able to be induced by the heavy right-handed neutrinos via the Dirac mass terms supplemented by the Higgs scalar. If the heavy right-handed neutrinos come in two families, all of CP-violating phases associated with the Dirac neutrino masses can be converted into the CP-violating Dirac and Majorana phases associated with the light flavor neutrino masses. The model with two extra heavy right-handed neutrinos is called minimal seesaw model \cite{MinimalSeesaw}. If the seesaw mechanism is the right answer to give tiny neutrino masses, our relation Eq.(\ref{Eq:M_nu for 23}) is also described by more fundamental quantities, namely the Dirac neutrino masses.

In this article, within the framework of the minimal seesaw model, we would like to argue how Dirac neutrino masses are constrained so as to induce maximal CP violation and simultaneously to induce maximal atmospheric neutrino mixing as well \cite{MassTextureCPEarlier,mu-tau-CP}. In Sec.\ref{sec:2}, we introduce six Dirac neutrino masses associated with two extra heavy right-handed neutrinos. Three relations determining three neutrino mixing angles such as Eq.(\ref{Eq:M_nu for 23}) are expressed in terms of these six Dirac neutrino masses and the CP-violating Dirac phase, which are used to find constraints to induce maximal CP violation. To obtain simple and useful relations in the minimal seesaw model, we choose one combination of Dirac neutrino masses to vanish, which includes texture one zero. The detailed discussions to reach various constraints on Dirac neutrino masses are presented in Appendix \ref{sec:Appendix}. In Sec.\ref{sec:3}, we derive necessary constraints on the Dirac neutrino masses to induce maximal CP violation. Finally, further assuming maximal atmospheric neutrino mixing, we determine six Dirac neutrino masses to be real or imaginary.  The final section Sec.\ref{sec:4} is devoted to summary and discussions, which include a preliminary argument on the creation of the baryon number of the universe via the leptogenesis based on our constraints on the Dirac neutrino masses.

%%%%%%%%%%%%%%%%%%%%%%%%%%%%%%%%%%%%%%%%%
\section{\label{sec:2}Dirac masses and Dirac CP violation}
%%%%%%%%%%%%%%%%%%%%%%%%%%%%%%%%%%%%%%%%%
The minimal seesaw model contains two extra right-handed neutrinos. We understand that a $2 \times 2$ heavy neutrino mass matrix $M_R$ and a charged lepton mass matrix are transformed into diagonal and real ones. After the heavy right-handed neutrinos are decoupled, the minimal seesaw mechanism generates a symmetric $3 \times 3$ light neutrino mass matrix $M_\nu$ containing $M_{ij}$ as elements to yield $M_\nu = -m_D M_R^{-1} m_D^T$, where $m_D$ is a $3 \times 2$ Dirac neutrino mass matrix. We parameterize $M_R$ by 
%%%%%%%%%%%%%%%%%%%%
\begin{eqnarray}
M_R = \left(
  \begin{array}{cc}
  M_1 & 0   \\ 
  0       & M_2
  \end{array}
\right)\quad (M_1<M_2),
\end{eqnarray}
%%%%%%%%%%%%%%%%%%%%
and $m_D$ by
%%%%%%%%%%%%%%%%%%%%
\begin{eqnarray}
m_D = 
\left(
  \begin{array}{cc}
    \sqrt{M_1}a_1  & \sqrt{M_2}b_1   \\
    \sqrt{M_1}a_2  & \sqrt{M_2}b_2   \\
    \sqrt{M_1}a_3  & \sqrt{M_2}b_3   \\
  \end{array}
\right),
\end{eqnarray}
%%%%%%%%%%%%%%%%%%%%
which result in
%%%%%%%%%%%%%%%%%%%%
\begin{eqnarray}
M_\nu &=& \left( \begin{array}{*{20}{c}}
M_{ee}&M_{e\mu }&M_{e\tau }\\
M_{e\mu }&M_{\mu \mu }&M_{\mu \tau }\\
M_{e\tau }&M_{\mu \tau }&M_{\tau \tau }
\end{array} \right)
=
-\left(
  \begin{array}{ccc}
    a_1^2 + b_1^2   & a_1a_2 + b_1b_2   &  a_1a_3 + b_1b_3  \\
    a_1a_2 + b_1b_2 & a_2^2 + b_2^2     &  a_2a_3 + b_2b_3  \\
    a_1a_3 + b_1b_3 & a_2a_3 + b_2b_3   &  a_3^2 + b_3^2   \\
  \end{array}
\right),
\label{Eq:Mnu}
\end{eqnarray}
%%%%%%%%%%%%%%%%%%%%
where the minus sign in front of the mass matrix is discarded for the later discussions.  One of the masses of $\nu_{1,2,3}$ is required to vanish owing to $\det\left(M_\nu\right)=0$. 

The useful relation Eq.(\ref{Eq:M_nu for 23}) expressed in terms of $M_{ij}$ is converted into
%%%%%%%%%%%%%%%%%%%%
\begin{eqnarray}
{a_ + }{a_ - } + {b_ + }{b_ - } =  - {t_{13}}\left( {{a_1}{a_ - } + {b_1}{b_ - }} \right){e^{ - i{\delta _{CP}}}},
\label{Eq:23 for seesaw} 
\end{eqnarray}
%%%%%%%%%%%%%%%%%%%%
where ${a_ + } = {s_{23}}{a_2} + {c_{23}}{a_3}$, ${b_ + } = {s_{23}}{b_2} + {c_{23}}{b_3}$, ${a_ - } = {c_{23}}{a_2} - {s_{23}}{a_3}$ and ${b_ - } = {c_{23}}{b_2} - {s_{23}}{b_3}$.  There are two more similar relations to Eq.(\ref{Eq:M_nu for 23}) that determine $\theta_{12,13}$ for given $M_{ij}$ \cite{Useful2} and these two relations give rise to    
%%%%%%%%%%%%%%%%%%%%
\begin{eqnarray}
&&
\sin 2{\theta _{12}}\left[ {\frac{{c_{13}^2\left( {a_1^2 + b_1^2} \right) - s_{13}^2\left( {a_ + ^2 + b_ + ^2} \right)}{e^{2i{\delta _{CP}}}}}{{\cos 2{\theta _{13}}}} - \left( {a_ - ^2 + b_ - ^2} \right)} \right] =  - 2\cos 2{\theta _{12}}\frac{{{a_1}{a_ - } + {b_1}{b_ - }}}{{{c_{13}}}},
\label{Eq:12 for seesaw}\\
&&
\sin 2{\theta _{13}}\left[ {\left( {a_ + ^2 + b_ + ^2} \right){e^{i{\delta _{CP}}}} - \left( {a_1^2 + b_1^2} \right){e^{ - i{\delta _{CP}}}}} \right] = 2\cos 2{\theta _{13}}\left( {{a_1}{a_ + } + {b_1}{b_ + }} \right).
\label{Eq:13 for seesaw}
\end{eqnarray}
%%%%%%%%%%%%%%%%%%%%
Similarly, neutrino masses accompanied by Majorana phases are calculated to be: 
%%%%%%%%%%%%%%%%%%%%
\begin{eqnarray}
{m_1}{e^{ - i{\phi _1}}} &=& a_ - ^2 + b_ - ^2 - \frac{{{a_1}{a_ - } + {b_1}{b_ - }}}{{{t_{12}}{c_{13}}}},
\nonumber\\
{m_2}{e^{ - i{\phi _2}}} &=& a_ - ^2 + b_ - ^2 + \frac{{{t_{12}}}}{{{c_{13}}}}\left( {{a_1}{a_ - } + {b_1}{b_ - }} \right),
\label{Eq:masses for seesaw}\\
{m_3}{e^{ - i{\phi _3}}} &=& \frac{{c_{13}^2\left( {a_ + ^2 + b_ + ^2} \right) - s_{13}^2\left( {a_1^2 + b_1^2} \right)}{e^{ - 2i{\delta _{CP}}}}}{{\cos 2{\theta _{13}}}}.
\nonumber
\end{eqnarray}
%%%%%%%%%%%%%%%%%%%%

These three relations Eqs.(\ref{Eq:23 for seesaw}), (\ref{Eq:12 for seesaw}) and (\ref{Eq:13 for seesaw}) can be casted into more compact forms since one of three neutrino masses turns out be zero owing to $\det(M_\nu)=0$. For NH, we have $m_1=0$ leading to
%%%%%%%%%%%%%%%%%%%%
\begin{equation}
a_ - ^2 + b_ - ^2 = \frac{1}{{{t_{12}}{c_{13}}}}\left( {{a_1}{a_ - } + {b_1}{b_ - }} \right),
\label{Eq:m1=0 for seesaw}
\end{equation}
%%%%%%%%%%%%%%%%%%%%
and obtain that
%%%%%%%%%%%%%%%%%%%%
\begin{eqnarray}
&&
{a_ + }{a_ - } + {b_ + }{b_ - } =  - {t_{13}}\left( {{a_1}{a_ - } + {b_1}{b_ - }} \right){e^{ - i{\delta _{CP}}}},
\label{Eq:23 for NHseesaw}\\
&&
a_1^2 + b_1^2 - t_{12}^2\left( {a_ - ^2 + b_ - ^2} \right) = {t_{13}}\left( {{a_1}{a_ + } + {b_1}{b_ + }} \right){e^{i{\delta _{CP}}}},
\label{Eq:12 for NHseesaw}\\
&&
c_{13}^2\left( {a_1^2 + b_1^2} \right) - s_{13}^2\left( {a_ + ^2 + b_ + ^2} \right){e^{2i{\delta _{CP}}}} = \left( {c_{13}^2 - s_{13}^2} \right)t_{12}^2\left( {a_ - ^2 + b_ - ^2} \right),
\label{Eq:13 for NHseesaw}
\end{eqnarray}
%%%%%%%%%%%%%%%%%%%%
and
%%%%%%%%%%%%%%%%%%%%
\begin{eqnarray}
{m_2}{e^{ - i{\phi _2}}} &=& \frac{1}{{c_{12}^2}}\left( {a_ - ^2 + b_ - ^2} \right),
\nonumber\\
{m_3}{e^{ - i{\phi _3}}} &=& \frac{1}{{c_{13}^2}}\left[ {a_ + ^2 + b_ + ^2 - s_{13}^2t_{12}^2\left( {a_ - ^2 + b_ - ^2} \right){e^{ - 2i{\delta _{CP}}}}} \right].
\label{Eq:masses for NHseesaw}
\end{eqnarray}
%%%%%%%%%%%%%%%%%%%%
For IH, we have $m_3=0$ leading to
%%%%%%%%%%%%%%%%%%%%
\begin{equation}
a_ + ^2 + b_ + ^2 = t_{13}^2\left( {a_1^2 + b_1^2} \right){e^{ - 2i{\delta _{CP}}}},
\label{Eq:m3=0 for seesaw}
\end{equation}
%%%%%%%%%%%%%%%%%%%%
and obtain that
%%%%%%%%%%%%%%%%%%%%
\begin{eqnarray}
&&
{a_ + }{a_ - } + {b_ + }{b_ - } =  - {t_{13}}\left( {{a_1}{a_ - } + {b_1}{b_ - }} \right){e^{ - i{\delta _{CP}}}},
\label{Eq:23 for IHseesaw}\\
&&
\sin 2{\theta _{12}}\left( {\frac{{a_1^2 + b_1^2}}{{c_{13}^2}} - \left( {a_ - ^2 + b_ - ^2} \right)} \right) =  - 2\cos 2{\theta _{12}}\frac{{{a_1}{a_ - } + {b_1}{b_ - }}}{{{c_{13}}}},
\label{Eq:12 for IHseesaw}\\
&&
{a_1}{a_ + } + {b_1}{b_ + } =  - {t_{13}}\left( {a_1^2 + b_1^2} \right){e^{ - i{\delta _{CP}}}},
\label{Eq:13 for IHseesaw}
\end{eqnarray}
%%%%%%%%%%%%%%%%%%%%
and
%%%%%%%%%%%%%%%%%%%%
\begin{eqnarray}
{m_1}{e^{ - i{\phi _1}}} &=& a_ - ^2 + b_ - ^2 - \frac{{{a_1}{a_ - } + {b_1}{b_ - }}}{{{t_{12}}{c_{13}}}},
\nonumber\\
{m_2}{e^{ - i{\phi _2}}} &=& a_ - ^2 + b_ - ^2 + \frac{{{t_{12}}}}{{{c_{13}}}}\left( {{a_1}{a_ - } + {b_1}{b_ - }} \right).
\label{Eq:masses for IHseesaw}
\end{eqnarray}
%%%%%%%%%%%%%%%%%%%%

We would like to obtain simple solutions to these equations for $a_{1,+,-}$ and $b_{1,+,-}$ and choose several plausible sets of the solutions, which are consistent with the hierarchical condition of $m^2_3\gg m^2_2$ requiring that $\left| a_ + ^2 + b_ + ^2 \right|^2\gg\left| a_ - ^2 + b_ - ^2 \right|^2$ for NH or $m^2_1\approx m^2_2$ requiring that $\left|a_ - ^2 + b_ - ^2 \right|^2\gg\left|{{a_1}{a_ - } + {b_1}{b_ - }} \right|^2$ for IH. As stated in the Introduction, we choose one combination of Dirac neutrino masses to vanish, which includes texture one zero. The discussions on our choices of the solutions are presented in Appendix \ref{sec:Appendix}, from which we can summarize our results as follows: For NH, 
%%%%%%%%%%%%%%%%%%%%
\begin{enumerate}
%%%%%%%%%%%%%%%%%%%%
\item in the case of $a_1 = 0$, $a_{+,-}$ and $b_{1,+,-}$ should satisfy ${a_ - } =  - {{{s_{13}}}}{a_ + }{e^{i{\delta _{CP}}}}/{{{t_{12}}}}$ and ${b_1} = {{{t_{12}}}}{b_ - }/{{{c_{13}}}} + {t_{13}}{b_ + }{e^{i{\delta _{CP}}}}$ as well as ${a_ + }{a_ - } =  - \left( {{b_ + } + {t_{13}}{b_1}e^{ - i{\delta _{CP}}}} \right){b_ - }$ and $a_ - ^2 + b_ - ^2 = {b_1}{b_ - }/{{{t_{12}}{c_{13}}}}$; %%%%%%%%%%%%%%%%%%%%
\item in the case of $b_1 = 0$, $a_{1,+,-}$ and $b_{+,-}$ should satisfy relations in the case of $a_1 = 0$ with the interchange of $a\leftrightarrow b$; 
%%%%%%%%%%%%%%%%%%%%
\item in the case of $a_- = 0$, $a_{1,+}$ and $b_{1,+,-}$ should satisfy ${a_1} = {t_{13}}{a_ + }{e^{i{\delta _{CP}}}}$, ${b_ - } = b_1/{t_{12}c_{13}}$ and ${b_ + } =  - {t_{13}}{b_1}{e^{ - i{\delta _{CP}}}}$;
%%%%%%%%%%%%%%%%%%%%
\item in the case of $b_- = 0$, $a_{1,+,-}$ and $b_{1,+}$ should satisfy relations in the case of $a_- = 0$ with the interchange of $a\leftrightarrow b$; 
%%%%%%%%%%%%%%%%%%%%
\item in the cases of $a_+ = 0$ and $b_+ = 0$, no simple linear expressions arise.
%%%%%%%%%%%%%%%%%%%%
\end{enumerate}
%%%%%%%%%%%%%%%%%%%%
and, for IH, we find that
%%%%%%%%%%%%%%%%%%%%
\begin{enumerate}
%%%%%%%%%%%%%%%%%%%%
\item in the case of ${{a_ + } = - {t_{13}}{a_1}{e^{ - i{\delta _{CP}}}}}$, $a_-$ and $b_{1,+,-}$ should satisfy ${b_ + } =  - {t_{13}}{b_1}{e^{ - i{\delta _{CP}}}}$;
%%%%%%%%%%%%%%%%%%%%
\item in the case of $a_1 = 0$, $a_{-,+}$ and $b_{1,+,-}$ should satisfy ${a_ + } = 0$ and ${b_ + } =  - {t_{13}}{b_1}{e^{ - i{\delta _{CP}}}}$;
%%%%%%%%%%%%%%%%%%%%
\item in the case of $b_1 = 0$, $a_{1,+,-}$ and $b_{+,-}$ should satisfy relations in the case of $a_1 = 0$ with the interchange of $a\leftrightarrow b$;
%%%%%%%%%%%%%%%%%%%%
\item the case of $a_+ = 0$ ($b_+ = 0$) is identical to the case 2 (the case 3);
%%%%%%%%%%%%%%%%%%%%
\item the case of $a_- = 0$ ($b_- = 0$) is included in the case 1 or 2 (the case 1 or 3) as an additional requirement.
%%%%%%%%%%%%%%%%%%%%
\end{enumerate}
%%%%%%%%%%%%%%%%%%%%
The case of 5 for NH is not further discussed because it does not supply no useful linear relations with respect $a_{1,+,-}$ and $b_{1,+,-}$ and the cases of 4 and 5 for IH are irrelevant.

%%%%%%%%%%%%%%%%%%%%%%%%%%%%%%%%%%%%%%%%%
\section{\label{sec:3}Maximal CP Violation}
%%%%%%%%%%%%%%%%%%%%%%%%%%%%%%%%%%%%%%%%%
In this section, we would like to find appropriate conditions on $a_{1,2,3}$ and $b_{1,2,3}$, which are similar to Eqs.(\ref{Eq:Maximal-CP1}) and (\ref{Eq:Maximal-CP2}), to induce maximal CP violation. From the discussions in Sec.\ref{sec:2}, we find several such candidates in both NH and IH.  We choose the phase to be ${e^{-i{\delta _{CP}}}}$ appearing in the equations as much the same way as in Eq.(\ref{Eq:M_nu for 23}). The results are summarized in TABLE \ref{Tab:NH} for NH and TABLE \ref{Tab:IH} for IH that show which Dirac neutrino masses are real or imaginary. In these tables, the real or imaginary Dirac neutrino masses give maximal CP violation through the relevant constraint(s).

%%%%%%%%%%%%%%%%%%%%
\begin{table}[htbp]
\begin{center}
\begin{tabular}{|c|c|c|c|}
\hline
 case&relevant constraint for $\delta _{CP}=\pm \pi/2$ &real&imaginary \\ \hline
 1& ${a_ - }{e^{-i{\delta _{CP}}}} =  - {{{s_{13}}}}{a_ + }/{{{t_{12}}}}$&$a_-$ &$a_+$\\ \hline
 2& ${b_ - }{e^{-i{\delta _{CP}}}} =  - {{{s_{13}}}}{b_ + }/{{{t_{12}}}}$&$b_-$ &$b_+$\\ \hline
 3& ${a_1}{e^{-i{\delta _{CP}}}} = {t_{13}}{a_ + }$, ${b_ + } =  - {t_{13}}{b_1}{e^{ - i{\delta _{CP}}}}$&$a_1$, $b_1$ &$a_+$, $b_+$\\ \cline{1-2}
 4& ${b_1}{e^{-i{\delta _{CP}}}} = {t_{13}}{b_ + }$, ${a_ + } =  - {t_{13}}{a_1}{e^{ - i{\delta _{CP}}}}$&&\\
\hline
\end{tabular}
\caption{\label{Tab:NH}Constraints for NH to induce maximal CP violation}
\end{center}
\end{table}

\begin{table}[htbp]
\begin{center}
\begin{tabular}{|c|c|c|c|}
\hline
 case&relevant constraint for $\delta _{CP}=\pm \pi/2$ & real & imaginary \\ \hline
 1& ${a_ + } =  - {t_{13}}{a_1}{e^{ - i{\delta _{CP}}}}$, ${b_ + } =  - {t_{13}}{b_1}{e^{ - i{\delta _{CP}}}}$&$a_1$, $b_1$ &$a_+$, $b_+$\\ \hline
 2& ${b_ + } =  - {t_{13}}{b_1}{e^{ - i{\delta _{CP}}}}$&$b_1$ &$b_+$\\ \hline
 3& ${a_ + } =  - {t_{13}}{a_1}{e^{ - i{\delta _{CP}}}}$&$a_1$ &$a_+$\\
\hline
\end{tabular}
\caption{\label{Tab:IH}Constraints for IH to induce maximal CP violation}
\end{center}
\end{table}
%%%%%%%%%%%%%%%%%%%%

If the atmospheric neutrino mixing is maximal as well, $a_+$ and $a_-$ turn out to be ${a_ + } = (\sigma{a_2} + {a_3})/{\sqrt 2}$ and ${a_ - } = \sigma (\sigma{a_2}- {a_3})/{\sqrt 2}$. Therefore, it can be observed that the relation of ${a_3} = -\sigma{a_2}^\ast$ as long as $a_+\neq 0$ and $a_-\neq 0$ ensures the appearance of the imaginary $a_+$ in all focused cases requiring $a_-$ to be real and similarly for $b_{+,-}$.  This constraint on $a_{2,3}$ (or $b_{2,3}$) is equivalent to Eqs.(\ref{Eq:Maximal-CP1}) and (\ref{Eq:Maximal-CP2 and ATM}). In terms of $a_{1,+,-}$ and $b_{1,+,-}$, Eqs.(\ref{Eq:Maximal-CP1}) and (\ref{Eq:Maximal-CP2 and ATM}) can be expressed as ${\rm Re} \left( {{a_ + }{a_ - } + {b_ + }{b_ - }} \right) = 0$ and ${\rm Im} \left( {{a_1}{a_ - } + {b_1}{b_ - }} \right) = 0$.

So far, we have assumed that one of $a_{1,+,-}$ and $b_{1,+,-}$ vanishes but more general conclusion can be obtained without making any assumptions.  It is known that the relations of $M_{e\tau} = -\sigma M_{e\mu}^\ast$ and $M_{\tau\tau} = M_{\mu\mu}^\ast$ supplemented by $M_{ee,\mu\tau}$=real lead to maximal CP violation as well as maximal atmospheric neutrino mixing \cite{MassTextureCPEarlier,MassTextureCP}. In our point of view, it is understood that these relations serve as specific solutions to Eqs.(\ref{Eq:Maximal-CP1}) and (\ref{Eq:Maximal-CP2 and ATM}) \cite{Useful1}.  In terms of $a_{1,+,-}$ and $b_{1,+,-}$, the solution consists of ${a_3} = -\sigma{a_2}^\ast$ and ${b_3} = -\sigma{b_2}^\ast$ supplemented by ${a_1} = {\rm real}$ and ${b_1} = {\rm real}$. The Dirac neutrino masses are uniquely determined to be:
%%%%%%%%%%%%%%%%%%%%
\begin{eqnarray}
m_D = 
\left(
  \begin{array}{cc}
    \sqrt{M_1}a_1  & \sqrt{M_2}b_1   \\
    \sqrt{M_1}a_2  & \sqrt{M_2}b_2   \\
    \sqrt{M_1}(-\sigma a^\ast_2)  & \sqrt{M_2}(-\sigma b^\ast_2)   \\
  \end{array}
\right),
\label{Eq:MaximalCPmD}
\end{eqnarray}
%%%%%%%%%%%%%%%%%%%%
where $a_1$ and $b_1$ are real. As in Ref.\cite{MassTextureCP,FlavoredSym}, if a unitary matrix $S$ is defined to be
%%%%%%%%%%%%%%%%%%%%
\begin{eqnarray}
S = \left( {\begin{array}{*{20}{c}}
1&0&0\\
0&0&{ - \sigma }\\
0&{ - \sigma }&0
\end{array}} \right),
\label{Eq:mu-tau}
\end{eqnarray}
%%%%%%%%%%%%%%%%%%%%
on the $\left(\nu_e,\nu_\mu,\nu_\tau\right)$ basis, which provides $\mu$-$\tau$ flavored CP symmetry for the flavor neutrinos \cite{FlavoredSym} subjected to the interchange of $\nu_\mu$ and $\nu_\tau$, it is found that $m_D$ of Eq.(\ref{Eq:MaximalCPmD}) satisfies that $S^Tm_D = m^\ast_D$ as expected. 

%%%%%%%%%%%%%%%%%%%%%%%%%%%%%%%%%%%%%%%%%
\section{\label{sec:4}Summary and Discussions}
%%%%%%%%%%%%%%%%%%%%%%%%%%%%%%%%%%%%%%%%%
We are able to derive the useful and simple relations to induce maximal CP violation, which dictate that $a_+$=imaginary and $a_1$=real and/or $b_+$=imaginary and $b_1$=real for both NH and IH. For NH, either $a_+$=imaginary and $a_-$=real or $b_+$=imaginary and $b_-$=real also arises. These relations are limited to hold in specific textures where at least one of $a_{1,+,-}$ and $b_{1,+,-}$ vanishes. If the atmospheric neutrino mixing is also maximal, we have obtained ${a_3} = -\sigma{a_2}^\ast$, ${b_3} = -\sigma{b_2}^\ast$, $a_1$=real and $b_1$=real applicable to more general textures. These relations turn out to be equivalent to the familiar relations of $M_{e\tau}=-\sigma M^\ast_{e\mu}$,$M_{\tau\tau}=M^\ast_{\mu\mu}$, $M_{ee}={\rm real}$ and $M_{\mu\tau}={\rm real}$ for flavor neutrinos.  

Our findings about various relations among the Dirac masses giving the maximal CP violation for flavor neutrinos become useful when neutrino physics is affected by phases of the Dirac masses. The immediate such an example is to apply our method to the process of the creation of the baryon number of the universe via the leptogenesis. In fact, the result indicates that the CP-violating Majorana phases for the leptogenesis come from $a_{2,3}$ and $b_{2,3}$ (see Eq.(\ref{Eq:Delta})) although there is no Majorana CP violation for flavor neutrinos if the above relations are satisfied. 

To see how the baryon-photon ratio in the universe via the leptogenesis scenario can be predicted by the use of our requirement on the Dirac neutrino masses for the maximal CP violation, we provide a preliminary result as a viable example. First of all, we summarize the recipes, which are known as follows \cite{leptogenesis,FlavoredLeptogenesis}: 
%%%%%%%%%%%%%%%%%%%
\begin{itemize}
%%%%%%%%%%%%%%%%%%%
\item The CP asymmetry parameters from the decay of the lightest right-handed neutrino $N_1$ (we assume $M_1 \ll M_2$) is obtained from 
\begin{eqnarray}
\epsilon_i= -\frac{3M_1}{16\pi v^2}\frac{{\rm Im}[a_i^\ast b_i(a_1^\ast b_1 + a_2^\ast b_2 + a_3^\ast b_3)]}{\vert a_1 \vert^2+\vert a_2 \vert^2+\vert a_3 \vert^2},
\end{eqnarray}
where $i=e,\mu,\tau=1,2,3$ and $v \simeq 174$ GeV. 
%%%%%%%%%%%%%%%%%%%
\item The baryon number in the co-moving volume is calculated to be
\begin{eqnarray}
Y_B \simeq -\frac{12}{37 g_\ast} \left [(\epsilon_e+\epsilon_\mu)\eta\left(\frac{417}{589}(\vert a_1\vert^2 + \vert a_2\vert^2 )\right) + \epsilon_\tau \eta\left(\frac{390}{589} \vert a_3 \vert^2\right)  \right],
\end{eqnarray}
for $10^9 \le M_1 {\rm [GeV]} \le 10^{12}$ where washout effect on $\epsilon_i$ in the expanding universe is controlled by  
\begin{eqnarray}
\eta(x) = \left( \frac{8.25\times 10^{-3} {\rm eV}}{x} + \left( \frac{x}{2\times 10^{-4} {\rm eV}} \right)^{1.16} \right)^{-1},
\end{eqnarray}
and $g_\ast$ denotes the effective number of relativistic degree of freedom. We take $g_\ast = 106.75$.
%%%%%%%%%%%%%%%%%%%
\item  The baryon-photon ratio $\eta_B$ is estimated to be $\eta_B=7.04Y_B$.
%%%%%%%%%%%%%%%%%%%
\end{itemize}
%%%%%%%%%%%%%%%%%%%
Next, we estimate the baryon-photon ratio by assuming the maximal CP violation and the maximal atmospheric neutrino mixing in the neutrino sector: e.g., $a_1={\rm real}$, $b_1={\rm real}$, $a_3=-\sigma a_2^\ast$ and $b_3=-\sigma b_2^\ast$. In this case, there are only two independent phases ${\rm arg}(a_2)$ and ${\rm arg}(b_2)$. The CP asymmetry parameter $\epsilon_i$ are obtained as
\begin{eqnarray}
\epsilon_e &=& 0,  \nonumber \\
\epsilon_\mu &=&  -\frac{3M_1}{16\pi v^2}\frac{(|a_1||b_1| + 2{\rm Re}[a_2^\ast b_2] ){\rm Im}[a_2^\ast b_2]}{|a_1|^2+2 \vert a_2 \vert^2} \nonumber \\
&=&  -\frac{3M_1}{16\pi v^2}\frac{(|a_1||b_1| + 2|a_2||b_2|\cos\Delta)|a_2||b_2|\sin\Delta}{|a_1|^2+2 |a_2|^2}, \nonumber \\
\epsilon_\tau &=& -\epsilon_\mu,
\label{Eq:epsilon_maxCP_maxAtm}
\end{eqnarray}
where 
%%%%%%%%%%%%%%%%%%%%
\begin{equation}
\Delta = {\rm arg}(b_2)-{\rm arg}(a_2). 
\label{Eq:Delta}
\end{equation}
%%%%%%%%%%%%%%%%%%%%
From Eq.(\ref{Eq:epsilon_maxCP_maxAtm}), as we expected, the phase difference $\Delta$ has crucial role in the baryon asymmetry generation in the universe and $\Delta \neq n\pi$ $(n=0, \pm 1, \pm 2 \cdots)$ is required for nonvanishing baryon-photon ratio. 

To confirm results of our discussions more concretely, we estimate the CP asymmetry parameters shown in Eq.(\ref{Eq:epsilon_maxCP_maxAtm}) with the horizontal equality in the Dirac mass matrix \cite{Barger2004PLB}. There are the following three cases of the horizontal equality for elements denoted by $X$: 
\begin{eqnarray}
{\rm I}:\left(
  \begin{array}{cc}
    X  & X\\
    * & *\\
    * & *\\
  \end{array}
\right),
\quad
{\rm II}:\left(
  \begin{array}{cc}
    *  & *\\
    X & X\\
    * & *\\
  \end{array}
\right),
\quad
{\rm III}:\left(
  \begin{array}{cc}
    * & *\\
    * & *\\
    X & X\\
  \end{array}
\right),
\end{eqnarray}
where the mark ``$*$" denotes a nonvanishing element. The vertical equality is also discussed \cite{He2011PRD}. In the case II and case III, we obtain $\Delta=0$. The case I only survives for the maximal CP violation as well as the maximal atmospheric neutrino mixing for nonvanishing baryon-photon ratio. The phenomenological consequences with the horizontal equality have been obtained by numerical calculations. In this paper, we show the clear constraint on the models with horizontal equality by exact analytical expressions. This is an advantage of our research.

We show a numerical example of the baryon-photon ratio in the case I of the horizontal equality requiring $\sqrt{M_1}a_1=\sqrt{M_2}b_1$ for the maximal CP violation and the maximal atmospheric neutrino mixing. The effective mass of the neutrino less double beta decay is estimated as $M_{ee}= (1+M_1/M_2)a_1^2$. For the sake of simplicity, we assume $|a_2|=|b_2|$ and $\Delta = \pi/2$. The CP asymmetry parameter $\epsilon_\mu$ is
\begin{eqnarray}
\epsilon_\mu = -\frac{3M_1}{16\pi v^2}\sqrt{\frac{M_1}{M_2}}\frac{|M_{ee}| |a_2|^2}{|M_{ee}|+2 (1+M_1/M_2) |a_2|^2},
\end{eqnarray}
and we obtain 
\begin{eqnarray}
\eta_B=6.1\times 10^{-10},
\end{eqnarray}
for $M_1=9.7\times 10^{11}$ GeV, $M_2=100M_1$, $|M_{ee}|=0.069$ eV and $|a_2|=0.063$ eV, which is consistent with the observed value of $\eta_B = (6.02-6.18) \times 10^{-10}$ \cite{etaB}. More general analysis will be found elsewhere \cite{New}.
%%%%%%%%%%%%%%%%%%%

\vspace{3mm}
%%%%%%%%%%%%%%%%%%%%%%%%%%%%%%%%%%%%%%%%%%%%%%%%%%%%%%%%%%%%%%%%%%%%%%%%%%%%%%%
%\noindent
%\centerline{\small \bf ACKNOWLEGMENTS}
%%%%%%%%%%%%%%%%%%%%%%%%%%%%%%%%%%%%%%%%%%%%%%%%%%%%%%%%%%%%%%%%%%%%%%%%%%%%%%%

%%%%%%%%%%%%%%%%%%%%%%%%%%%%%%%%%%%%%%%%%%%%%%%%%%%%%%%%%%%%%%%%%%%%%%%%%%%%%%%%% The Appendices part is started with the command \appendix;
\appendix

\section{\label{sec:Appendix}Useful Constraints}

In this appendix, we describe how to obtain various constraints on $a_{1,+,-}$ and $b_{1,+,-}$ as solutions to the equations, Eqs.(\ref{Eq:m1=0 for seesaw})-(\ref{Eq:13 for NHseesaw}) for NH and Eqs.(\ref{Eq:m3=0 for seesaw})-(\ref{Eq:13 for IHseesaw}) for IH. We use constraints on $a_{1,+,-}$ as initial conditions to find our solutions, which can be transformed into other solutions based on those on $b_{1,+,-}$ by the interchange of $a\leftrightarrow b$. The initial setup for $a_{1,+,-}$, where one combination of Dirac neutrino masses to vanish, turns out to be given by $a_1=0$, $a_+=0$, $a_-=0$ or ${{a_ + } + {t_{13}}{a_1}{e^{ - i{\delta _{CP}}}} = 0}$.  For NH,
%%%%%%%%%%%%%%%%%%%
\begin{enumerate}
\item $a_1=0$: From Eq.(\ref{Eq:m1=0 for seesaw}), ${a_ - ^2 =  - \left( {{b_ - } - \frac{1}{{{t_{12}}{c_{13}}}}{b_1}} \right){b_ - }}$ is required to have $m_1 = 0$. From Eq.(\ref{Eq:23 for NHseesaw}) for $\theta_{23}$ and Eq.(\ref{Eq:12 for NHseesaw}) for $\theta_{12}$, we, respectively, obtain ${a_ + }{a_ - } =  - \left( {{b_ + } + {t_{13}}{e^{ - i{\delta _{CP}}}}{b_1}} \right){b_ - }$ and ${b_1} = \frac{{{t_{12}}}}{{{c_{13}}}}{b_ - } + {t_{13}}{b_ + }{e^{i{\delta _{CP}}}}$, which turn out to satisfy (\ref{Eq:13 for NHseesaw}) for $\theta_{13}$. 
Inserting the expression of $b_1$ into those of $a_ - ^2$ and $a_+a_-$, finally, gives 
a simpler relation ${a_ - } =  - \frac{{{s_{13}}}}{{{t_{12}}}}{a_ + }{e^{i{\delta _{CP}}}}$.   We obtain that ${b_1} = \frac{{{t_{12}}}}{{{c_{13}}}}{b_ - } + {t_{13}}{b_ + }{e^{i{\delta _{CP}}}}$ and ${a_ - } =  - \frac{{{s_{13}}}}{{{t_{12}}}}{a_ + }{e^{i{\delta _{CP}}}}$  as useful relations together with $a_-=0$.
\item $a_+=0$: It is readily recognized that no simple linear relations are deduced from the equations and $a_{1,-}$ and $b_{1,+,-}$ should satisfy $a_ - ^2 + b_ - ^2 = \frac{1}{{{t_{12}}{c_{13}}}}\left( {{a_1}{a_ - } + {b_1}{b_ - }} \right)$  from Eq.(\ref{Eq:m1=0 for seesaw}), ${b_ + }{b_ - } =  - {t_{13}}{e^{ - i{\delta _{CP}}}}\left( {{a_1}{a_ - } + {b_1}{b_ - }} \right)$ from Eq.(\ref{Eq:23 for NHseesaw}), $a_1^2 + b_1^2 - t_{12}^2\left( {a_ - ^2 + b_ - ^2} \right) = {t_{13}}{b_1}{b_ + }{e^{i{\delta _{CP}}}}$ from Eq.(\ref{Eq:12 for NHseesaw}) and $c_{13}^2\left( {a_1^2 + b_1^2} \right) - s_{13}^2{e^{2i{\delta _{CP}}}}b_ + ^2 = \left( {c_{13}^2 - s_{13}^2} \right)t_{12}^2\left( {a_ - ^2 + b_ - ^2} \right)$ from Eq.(\ref{Eq:13 for NHseesaw}).
\item $a_-=0$: From Eq.(\ref{Eq:m1=0 for seesaw}), ${b_ - } = \frac{1}{{{t_{12}}{c_{13}}}}{b_1}$ is required to have $m_1 = 0$. From Eq.(\ref{Eq:23 for NHseesaw}) and Eq.(\ref{Eq:12 for NHseesaw}), we, respectively, obtain ${b_ + } =  - {t_{13}}{b_1}{e^{ - i{\delta _{CP}}}}$ and ${a_1} = {t_{13}}{a_ + }e^{i{\delta _{CP}}}$, which turn out to satisfy (\ref{Eq:13 for NHseesaw}) for $\theta_{13}$. We obtain that ${b_ + } + {t_{13}}{b_1}{e^{ - i{\delta _{CP}}}}=0$, ${b_ - } = \frac{1}{{{t_{12}}{c_{13}}}}{b_1}$ and ${a_1} = {t_{13}}{a_ + }e^{i{\delta _{CP}}}$ together with $a_1=0$.
\end{enumerate}
%%%%%%%%%%%%%%%%%%%
For IH, the combined use of Eqs.(\ref{Eq:m3=0 for seesaw}) and (\ref{Eq:13 for IHseesaw}) yields ${a_1}{a_ + } + {b_1}{b_ + } = -{t_{13}}\left( {a_1^2 + b_1^2} \right){e^{ - i{\delta _{CP}}}}$ for Eq.(\ref{Eq:13 for IHseesaw}) giving 
%%%%%%%%%%%%%%%%%%%%
\begin{equation}
\left( {{a_ + } + {t_{13}}{a_1}{e^{ - i{\delta _{CP}}}}} \right){a_1} + \left( {{b_ + } + {t_{13}}{b_1}{e^{ - i{\delta _{CP}}}}} \right){b_1} = 0,
\label{Eq:13-simplerIH}
\end{equation}
%%%%%%%%%%%%%%%%%%%%
by which Eq.(\ref{Eq:m3=0 for seesaw}) is further reduced to 
%%%%%%%%%%%%%%%%%%%%
\begin{equation}
\left( {{a_ + } + {t_{13}}{a_1}{e^{ - i{\delta _{CP}}}}} \right){a_ + } + \left( {{b_ + } + {t_{13}}{b_1}{e^{ - i{\delta _{CP}}}}} \right){b_ + } = 0.
\label{Eq:m3=0-simplerIH}
\end{equation}
%%%%%%%%%%%%%%%%%%%%
Similarly, Eq.(\ref{Eq:23 for IHseesaw}) leads to
%%%%%%%%%%%%%%%%%%%%
\begin{equation}
\left( {{a_ + } + {t_{13}}{a_1}{e^{ - i{\delta _{CP}}}}} \right){a_ - } + \left( {{b_ + } + {t_{13}}{b_1}{e^{ - i{\delta _{CP}}}}} \right){b_ - } = 0.
\label{Eq:23-simplerIH}
\end{equation}
%%%%%%%%%%%%%%%%%%%
Considering Eqs.(\ref{Eq:13-simplerIH})-(\ref{Eq:23-simplerIH}), we find the following cases:
\begin{enumerate}
\item ${\frac{{{b_ + } + {t_{13}}{b_1}{e^{ - i{\delta _{CP}}}}}}{{{a_ + } + {t_{13}}{a_1}{e^{ - i{\delta _{CP}}}}}} =  - \frac{{{a_1}}}{{{b_1}}} =  - \frac{{{a_ + }}}{{{b_ + }}} =  - \frac{{{a_ - }}}{{{b_ - }}}}$ for ${{a_ + } + {t_{13}}{a_1}{e^{ - i{\delta _{CP}}}}}\neq 0$, ${{b_ + } + {t_{13}}{b_1}{e^{ - i{\delta _{CP}}}}}\neq 0$ as well as $a_{1,+,-}\neq 0$ and  $b_{1,+,-}\neq 0$: Eq.(\ref{Eq:23-simplerIH}) with ${a_ + } = {b_ + }{a_1}/{b_1}$ yields either ${b_ + } + {t_{13}}{b_1}{e^{ - i{\delta _{CP}}}} = 0$, which is not allowed by the initial conditions, or ${a_1}{a_ - } + {b_1}{b_ - } = 0$ giving $\left| {{m_1}} \right| = \left| {{m_2}} \right|$ from Eq.(\ref{Eq:masses for IHseesaw}), which contradicts the fact that $\left| {{m_1}} \right| < \left| {{m_2}} \right|$. This case cannot provide a solution.
\item ${{a_ + } + {t_{13}}{a_1}{e^{ - i{\delta _{CP}}}} = 0}$: It is readily found that ${{b_ + } + {t_{13}}{b_1}{e^{ - i{\delta _{CP}}}} = 0}$ is the solution. We obtain that ${{a_ + } + {t_{13}}{a_1}{e^{ - i{\delta _{CP}}}} = 0}$ and ${{b_ + } + {t_{13}}{b_1}{e^{ - i{\delta _{CP}}}} = 0}$.
\item $a_1 = 0$: $\left( {{b_ + } + {t_{13}}{b_1}{e^{ - i{\delta _{CP}}}}} \right){b_ 1 } = 0$ is required and ${{b_ + } + {t_{13}}{b_1}{e^{ - i{\delta _{CP}}}}} = 0$ is the solution because ${b_ 1 } = 0$ gives $\left| {{m_1}} \right| = \left| {{m_2}} \right|$. The remaining conditions from  Eqs.(\ref{Eq:13-simplerIH})-(\ref{Eq:23-simplerIH}) are fulfilled by $a_+=0$. For Eq.(\ref{Eq:12 for IHseesaw}), $\sin 2{\theta _{12}}\left( {b_1^2 - c_{13}^2\left( {a_ - ^2 + b_ - ^2} \right)} \right) =  - 2\cos 2{\theta _{12}}{c_{13}}{b_1}{b_ - }$ should be satisfied. We obtain that ${{b_ + } + {t_{13}}{b_1}{e^{ - i{\delta _{CP}}}}} = 0$ and $a_1 = a_+=0$.
\item $a_ + = 0$: Either ${{b_ + } + {t_{13}}{b_1}{e^{ - i{\delta _{CP}}}} = 0}$ or $b_+=0$ is the solution. If $b_+=0$, Eq.(\ref{Eq:23-simplerIH}) yields ${{a_1}{a_ - } + {b_1}{b_ - } = 0}$, which results in ${\left| {{m_1}} \right| = \left| {{m_2}} \right|}$ from Eq.(\ref{Eq:masses for IHseesaw}). For ${{b_ + } + {t_{13}}{b_1}{e^{ - i{\delta _{CP}}}} = 0}$, $a_1=0$ is derived.  We obtain that ${{b_ + } + {t_{13}}{b_1}{e^{ - i{\delta _{CP}}}} = 0}$ and $a_ 1 = a_+=0$.
\item $a_ - = 0$: $\left( {{b_ + } + {t_{13}}{b_1}{e^{ - i{\delta _{CP}}}}} \right){b_ - } = 0$ is required and ${{b_ + } + {t_{13}}{b_1}{e^{ - i{\delta _{CP}}}}} = 0$ is the solution because ${b_ - } = 0$ gives $\left| {{m_1}} \right| = \left| {{m_2}} \right|$ from Eq.(\ref{Eq:masses for IHseesaw}). The remaining conditions are fulfilled by either ${a_ + } + {t_{13}}{a_1}{e^{ - i{\delta _{CP}}}} = 0$ or ${a_1} = {a_ + } = 0$. We obtain that ${a_ + } + {t_{13}}{a_1}{e^{ - i{\delta _{CP}}}} = 0$, ${{b_ + } + {t_{13}}{b_1}{e^{ - i{\delta _{CP}}}}} = 0$ and $a_ - = 0$ or that ${{b_ + } + {t_{13}}{b_1}{e^{ - i{\delta _{CP}}}}} = 0$ and ${a_1} = {a_ + } =a_ - = 0$.
\end{enumerate}
All of the cases for IH are not independent.  For instance, the case 5 is included in the case 2 or in the case 3 both with the additional condition of $a_-=0$.  
%%%%%%%%%%%%%%%%%%%

%% References

%%References
%%--------------------------------
%\bibliographystyle{elsarticle-num}



\end{document}